# Nanodevice-Enabled Near-Field Thermal Radiation between Sub-Wavelength Surfaces

Xiao Luo[1,+], Hakan Salihoglu[1,+], Zexiao Wang[1,+], Zhuo Li[1,+], Hyeonggyun Kim[1], Jiayu Li[1], Bowen Yu[1], Shen Du[1], Sheng Shen[1,*].

1 Department of Mechanical Engineering, Carnegie Mellon University, Pittsburgh, PA 15213, USA.

+ The authors contributed equally.

* Corresponding author: Sheng Shen, email: sshen1@cmu.edu

## Abstract

With the continuous advancement of nanotechnology, nanodevices have become crucial components in computing, sensing and energy conversion applications. However, the structures of nanodevices typically possess sub-wavelength dimensions and separations, which pose significant challenges for understanding energy transport phenomena in nanodevices. Here, based on a judiciously designed thermal nanodevice, we report the first measurement of near-field energy transport between two coplanar sub-wavelength structures over temperature bias up to ~190 K. Our experimental results demonstrate a remarkable 20-fold enhancement in heat transfer beyond blackbody radiation. In contrast with the well-established near-field interactions between two semi-infinite bodies, the sub-wavelength confinements in nanodevices lead to the increased polariton scattering and the reduction of supporting modes and therefore a lower heat flow at a given separation. Our work unveils exciting opportunities for the rational design of nanodevices, particularly for on-chip near-field energy transport, with important implications for the development of efficient nanodevices for energy harvesting and thermal management.

Nanodevices play a vital role in a broad range of emerging technologies including quantum computing[1], nanophotonic biosensors[2], and widely tunable nanolasers[3]. These devices often consist of active elements with sub-wavelength dimensions and separations ($<\lambda\sim$10 μm at room temperature), which give rise to unusual energy transport paths, such as super-Planckian far field thermal radiation[4], and quantum fluctuational energy transport between the sub-wavelength elements across vacuum separations via Casimir effect[5–7]. Particularly, at sub-wavelength separations, near-field thermal radiation exhibits remarkable energy transport enhancements, exceeding well-established blackbody radiation with orders of magnitudes[8–11]. Leveraging such near-field enhancements between nanoscale elements opens up extraordinary applications of nanodevices in thermal management[12,13] and energy harvesting[14–16]. However, the realization of near-field based thermal nanodevices has remained elusive due to the challenges in obtaining coplanar sub-wavelength elements and the lack of physical understanding for near-field thermal radiation between nanoscale elements. Current knowledge of near-field thermal radiation heavily relies on fluctuational electrodynamic framework derived based on semi-infinite[17,18] or multilayered structures[19–21], which assume infinite dimensions and neglect geometrical constrains of the surfaces. Consequently, previous experimental near-field studies[11,22–37] have extensively focused on the near-field radiation involving macroscopic structures much larger than wavelengths. In contrast, the use of sub-wavelength structures enables distinct energy transport physics in the near field, where lateral confinements[38] may support new guided surface modes, or filter certain radiation modes.

To unlock the potential of near-field thermal radiation in nanodevices, here we employ custom-built on-chip nanodevices, which is designed to measure the near-field thermal radiation between two coplanar sub-wavelength membranes with nanofabricated gaps ranging from ~150 nm to ~750 nm. By integrating with suspended platinum resistive heating and thermometry, our on-chip devices allow for precise detection of near-field radiative heat transfer between sub-wavelength surfaces over a wide range of temperature differences up to ~ 190 K. Due to the mode suppression from sub-wavelength confinement, the measured near-field heat flows are observed to be smaller than the analytical predictions derived for the well-known semi-infinite media. We also find that this discrepancy diminishes as the gaps decrease, especially when the gap is less than the smallest lateral dimension.

## On-chip near-field nanodevices

As depicted in Fig. 1a, our on-chip nanodevice comprises two suspended coplanar silicon nitride membranes with lateral dimensions of 300 nm in thickness ($t$) and 7 μm in width ($w$). The emitting and absorbing membranes are extruded from their corresponding bases by lengths ($l$) of 3.5 μm and 2 μm, respectively. A platinum serpentine heater fabricated on the emitting membrane is used to increase the membrane temperature, whereas a platinum thermometry sensor on the absorbing membrane measures the temperature rise induced by near-field radiative heat transfer. To avoid the influence of heat conduction, the silicon substrate beneath the silicon nitride membranes is etched to form a suspended region (marked in Fig. 1b). On the absorbing membrane, a 100-μm-long trench is cut behind the platinum to fabricate a 1D long beam structure as the resistive thermometry sensor for heat flow probing. To investigate the coupled near-field interactions between two sub-wavelength structures (protruded parts on the emitting and absorbing membranes), we fabricate the on-chip devices with a range of nanoscale separations from ~150 to ~750 nm using electron beam lithography. As shown in Fig. 1c for the tilted-angle scanning electron microscope (SEM) image of a typical nanodevice, the width of the suspended region is controlled to < 15 μm to ensure good coplanarity and minimal relative deflection of the membranes. The height profile measurements with a Zyro non-contact profilometer reveal a height deviation of ~50 nm across the separation gap (Fig. 1c inset). By designing the serpentine heater, we achieve a uniform temperature profile along y-axis within the first 2-μm-deep region and minimize the temperature increase at the base of the emitting membrane along x-axis (See Extended Data Fig. 2 for the validated temperature profile from finite element simulation). We specifically design for a uniform temperature distribution along the first 2 μm from the edge because the near-field radiation is dominated by the first ~ 2 μm from the edge (See Supplementary Information for near-field thermal emission from the emitting membrane). The suspended base efficiently dissipates heat along the y-axis due to the short pathway to the intact silicon wafer (as the heat sink), and the temperature drops to ambient temperature within a short distance, ensuring the minimized thermal emission from the base of the emitting membrane to the absorbing membrane. Meanwhile, we also design reference devices to measure any background radiation absorbed by the sensor beam. (See Method and Extended Data for design and measurement of reference devices).

## High-sensitivity near-field thermal measurements

Figure 2a illustrates the measurement scheme and the corresponding thermal circuit in our experiments. All measurements are performed under high vacuum ($10^{-3}$ Pa) with the ambient temperature maintained at 300 K. A DC current *I* is supplied to the serpentine heater for heating the emitting membrane. The variation of the heater temperature $\Delta T_h$ ($\Delta T_h = T_h - T_{amb}$) is monitored by supplying a small sinusoidal current (0.4 μA at the 211 Hz frequency) and measuring the corresponding voltage change, $\Delta V_h$ (See Supplementary Information for the calibration of the platinum thermometer and corresponding temperature determination). To measure the temperature variation of the absorbing membrane $\Delta T_s$ ($\Delta T_s = T_s - T_{amb}$), we use a Wheatstone bridge circuit[39] to detect the voltage change across the long beam platinum thermometry sensor, which dissipates heat to both ends via heat conduction. The temperature profile along the beam sensor results in the voltage change, $\Delta V_s$ (See Supplementary Information for the temperature profile along the beam sensor). Figure 2b shows the measured $\Delta T_h$ and $\Delta T_s$ for a 150 nm-gap device, while the inset demonstrates the corresponding readings of $\Delta V_h$ and $\Delta V_s$ for the supplied heating current. The corresponding thermal radiation, $Q_{rad}$, is determined by $Q_{rad} = G_s \Delta T_s$. (See Supplementary Information for the calibration of the heat conductance of the sensor $G_s$). This measurement procedure is repeated for all the devices with various separation gaps and reference devices over a range of ~190 K temperature bias ($\Delta T = \Delta T_h - \Delta T_s$). To determine the near-field thermal radiation between two extruded membranes $Q_{mem}$, we deduct the background radiation heat flow measured by the reference device $Q_{ref}$ from the measured heat flow $Q_{rad}$.

We measure the radiation heat flows $Q_{mem}$ between the extruded membranes in 12 on-chip nanodevices, categorized into 4 groups based on gap distances. Figure 2c shows the typical $Q_{mem}$ as a function of the temperature bias $\Delta T$ from each group of gap distances, in which the corresponding gap distances are measured from SEM images (See Extended Data Fig. 4 for measured radiation heat flows of all devices). Across all devices, we observe that the radiation heat flow $Q_{mem}$ increases with an increase in temperature bias. Furthermore, at a given temperature bias, $Q_{mem}$ rises with reduced gap separation due to the near-field effect. For instance, when the maximum temperature bias of 190 K is applied in our experiments, reducing the gap distance from 726 nm to 151 nm leads to a significant heat flow increase from ~4 nW to ~50 nW. This behavior confirms the enhanced near-field thermal radiation between the membranes at smaller gap distances.

To quantify the near-field enhancement of the membranes, we define an enhancement ratio, $\varepsilon = Q_{mem}/Q_{bb}$. Here, $Q_{bb} = \sigma A F_{12}(T_h^4 - T_s^4)$ is the radiation heat flow between two blackbodies with the same configuration as the on-chip device, where $\sigma$ is Stefan-Boltzmann constant, $A$ is exchange surface area and $F_{12}$ is the view factor between the exchange surfaces. We plot the enhancement ratio $\varepsilon$ as a function of temperature bias in Fig. 2d (See Extended Data Fig. 5 for enhancement ratio $\varepsilon$ of all devices). Interestingly, over the 190 K temperature bias range, we observe that $\varepsilon$ decreases as the temperature bias increases, which will be further discussed with the simulated thermal radiation spectrum. As gap distance increases from 151 nm to 726 nm, the maximum radiation heat flow decreases by 92% (~50 nW to ~4 nW), while the maximum enhancement ratio ($\varepsilon$) only decreases by 80% (~20 to ~4), because the view factor $F_{12}$ also drops significantly from ~0.6 to ~0.2 (See Supplementary Information for the calculation of view factor).

## Fluctuational-electromagnetic simulation and theory

To validate our near-field measurements, we conduct fluctuational-electromagnetic (fluctuational-EM) simulation using the fluctuational surface current (FSC) method[40,41], which considers the two coplanar silicon nitride membranes with the design dimensions of 7 μm ×2 μm ×0.3 μm. Figure 3a shows the surface mesh used for the FSC method, based on which the reduced heat flow spectra $S(\omega)$ are calculated. From this, the simulated near-field heat flow $Q_{sim}$ can be determined by:

$$Q_{sim} = \int (\Theta(\omega, T_h) - \Theta(\omega, T_s))S(\omega)d\omega. \tag{1}$$

where $\Theta(\omega, T) = \hbar\omega/(\exp(\hbar\omega/k_B T) - 1)$ represents the Plank energy per oscillator at the temperature $T$. In Fig. 3b, we compare the simulation results with the measurements from the four devices in Fig. 2c as a function of temperature bias. For all measured 12 devices, we further compare the simulation results with their measured radiation heat flows as a function of gap distance at multiple temperature bias in Fig. 3c. In both temperature- and gap-dependent comparisons, the measured heat flows exhibit good agreement with the simulation results for all the measured separation gaps, thus further validating our experimental results. To elaborate on the mechanism of near-field enhancement, we plot the reduced heat flow spectra of different gap distances in Fig. 3d. The major peak around 1.8×10¹⁴ rad/s corresponds to the resonance of surface phonon polaritons supported by the silicon nitride membrane (see Supplementary Information for the permittivity of silicon nitride). In Fig. 3d, the reduced heat flow dramatically increases with decreased gap distance, indicating the strong near-field effect. While the resonance frequency of surface phonon polaritons remains fixed with temperature (see Supplementary Information for reflection measurements from the silicon nitride for broad temperature range), the radiative thermal energy peak (based on Wien's law) shifts to higher frequencies as temperature increases, resulting in $Q_{bb}$ increases with a rate higher than that of $Q_{mem}$ for the same $\Delta T$. Consequently, the near-field enhancement $\varepsilon$ decreases with increasing $\Delta T$ as shown in Fig. 2d.

To elucidate the impact of sub-wavelength dimensions in near-field radiative heat transfer, we compare our simulation results for the sub-wavelength surfaces with the analytical predictions based on the near-field heat flux of two semi-infinite plates. Figure 4a illustrates that the sub-wavelength surfaces (solid lines) exchange less heat than the semi-infinite surfaces (dashed lines) for the same cross section area of $w \times t = 7 \times 0.3 \ \mu m^2$. The observed lower heat flow arises from the boundary scattering of surface phonon polaritons due to the presence of the sub-wavelength structures, in which some scattered waves escape from the gap and do not contribute to the near-field heat transfer.

In Fig. 4b, we define and plot a heat flow ratio, $\varphi_{sim} = Q_{sim}/Q_{semi}$ with respect to separation gap within the temperature range of interest, where $Q_{semi}$ is the predicted heat flow by multiplying the near-field heat flux between two semi-infinite plates with the same cross section area ($w \times t$). As a comparison, we also introduce the other heat flow ratio, $\varphi_{exp} = Q_{mem}/Q_{semi}$, which shows excellent agreement with $\varphi_{sim}$ in Fig. 4b. At large separations ($d > 500$ nm), the difference between $Q_{sim}$ (or $Q_{exp}$) and $Q_{semi}$ is significant (e.g., $\varphi_{sim} \sim 0.2$ at $d = 800$ nm). However, as the separation decreases, the heat transfer rate between the sub-wavelength surfaces increases and approaches that between the semi-infinite plates (e.g., $\varphi_{sim} > 0.5$ for $d < 300$ nm). The increase in $\varphi_{sim}$ (or $\varphi_{exp}$) depends on two main criteria. First, at large separations, the surface phonon polaritons have wavelengths, $\lambda$, which are larger than but comparable to $t$. At shorter separations, the surface phonon polaritons possess smaller wavelengths and interact less with the lateral dimensions (See Supplementary Information for polariton and wavelength relations). As a result, the scattering of these polaritons at shorter gap distances are less pronounced compared to that at larger separations. Figures 4c and d show the electric field profile around the hot region for separation gaps of 150 nm and 700 nm, respectively. As seen, the field away from the surface (light blue and cyan regions) is spatially distributed along the thickness due to scattering. However, the field closer to the surface (yellow and green regions) is more uniformly distributed, resembling the semi-infinite case (no lateral dependence of the field profile). This behavior leads to the heat transfer rate closer to that between the semi-infinite media at shorter distances. Second, the rate of the increase in $\varphi_{sim}$ (or $\varphi_{exp}$) enhances for the separations below the thickness of the membrane. Spatial confinement along the thickness prevents existence of polaritons with wavelengths larger than thickness. These polaritons are only supported along the width direction and become main contributors to energy transport at large separations. At separations less than the thickness, polaritons with wavelength smaller than thickness mainly transport energy across the gap, and these polaritons exist along both the thickness and the width. Thus, at the short separations, the contributing polaritons supported in both directions carry a higher total energy and approach the semi-infinite model, where polaritons exist in all directions without any spatial constraint.

## Conclusion

In conclusion, we design and fabricate on-chip nanodevices to measure the near-field thermal radiation between sub-wavelength co-planar membranes. Through high-sensitivity measurements, we accurately quantify the near-field heat flow, which shows excellent agreement with fluctuational electrodynamics simulations. Compared to blackbody radiation, we observe 20-fold enhancement in thermal radiation between the two sub-wavelength structures with a separation gap of ~150 nm. In addition, our analysis finds that the near-field thermal radiation between the sub-wavelength co-planar membranes remains lower than that based on the semi-infinite model, which is attributed to energy escaping from the sub-wavelength confinements due to scattering of polaritons and less contributions from polariton modes with large wavelengths. Our findings elucidate the near-field energy transfer between sub-wavelength structures, paving the way for the development of on-chip near-field devices for energy harvesting and thermal management.

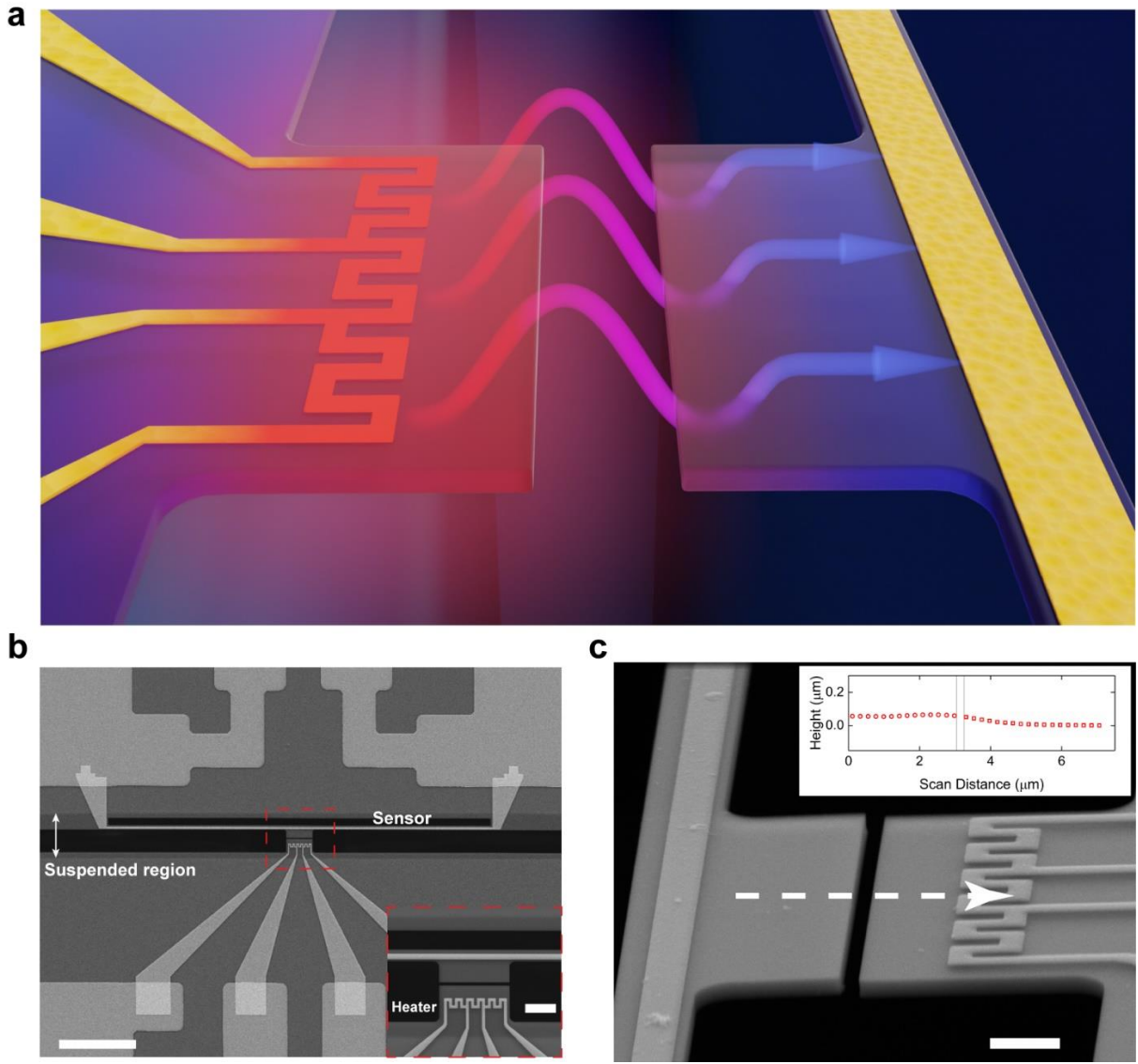

**Figure 1. Nanodevice enabled near-field thermal radiation between parallel sub-wavelength membranes**. **a**. Schematic of a near-field thermal nanodevice. The extruded emitting membrane (left) is heated up by platinum serpentine heater, while the extruded absorbing membrane (right) absorbs the near-field radiation with a temperature rise monitored by the platinum resistance thermometer. **b**. Top-view SEM images of the measurement nanodevice. Scale bar: 25 μm. The inset shows the zoomed-in SEM image. Scale bar: 3 μm. To physically isolate the emitting and absorbing membranes and thus exclude heat conduction, the heater and sensor structures are suspended by etching the beneath silicon substrate, as shown by the suspended region. The width of the suspended region is confined within 15 μm to maintain the coplanarity of the two membranes. **c**. 70°-tilted SEM image. Scale bar: 1 μm. The inset shows the measured profile along the dashed arrow indicating a height difference of ~50 nm.

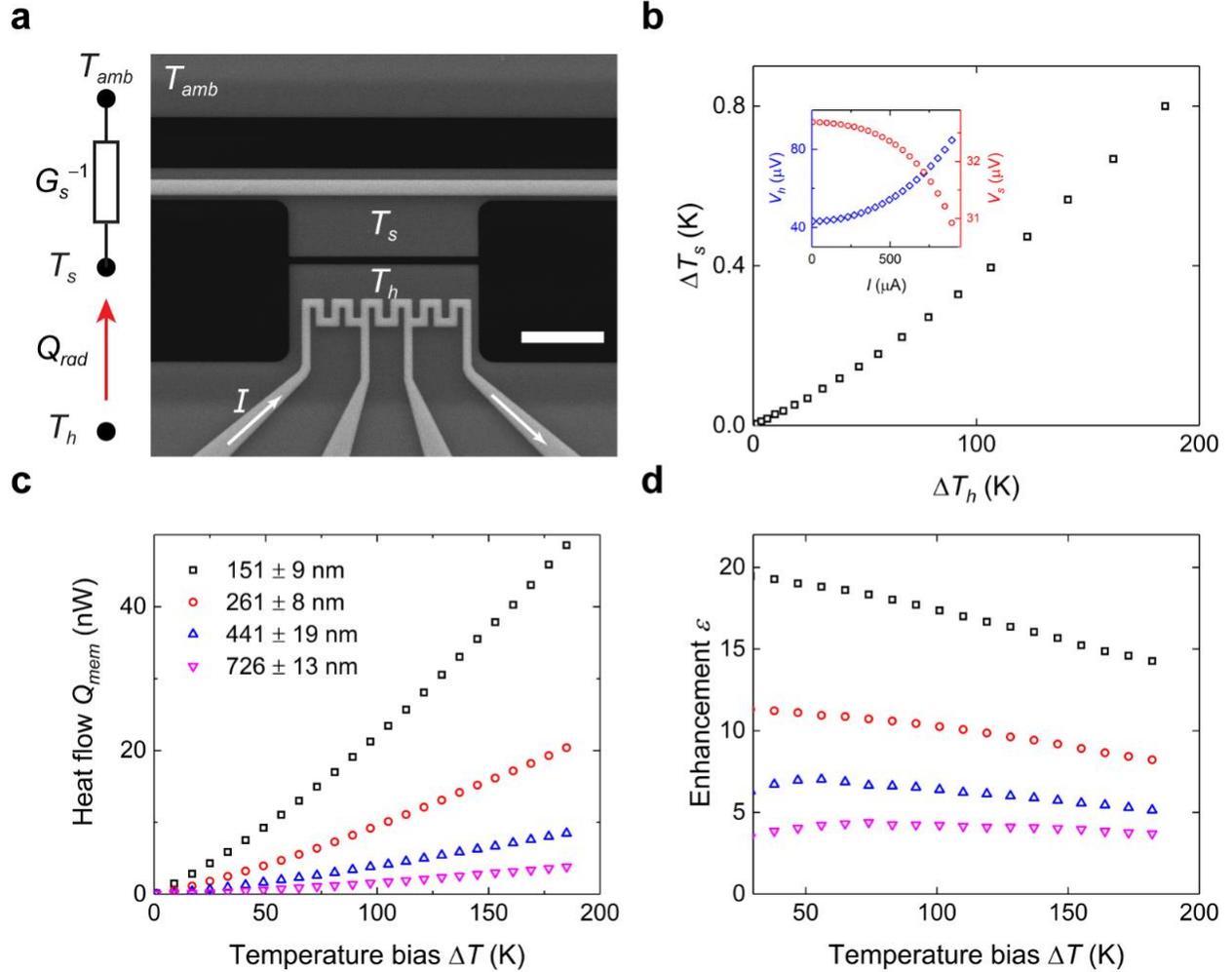

**Figure 2. Near-field measurement setup and experimental results. a**. Measurement setup and corresponding thermal circuit. The absorbed radiation heat flow $Q_{rad}$ is dissipated to the ambient via heat conduction, which induces a temperature rise of the sensor. Scale bar: 3 μm. **b**. Measured temperature increase of the absorbing membrane and corresponding temperature rise of the emitting membrane. The temperature increases of the emitting, $\Delta T_h$ and absorbing $\Delta T_s$ membranes are monitored by the four-probe voltage $V_h$ and gap voltage of the Wheatstone bridge $V_s$, respectively. The inset shows the measured voltages as a function of heating current, $I$, for a near-field device with a 151 nm separation. **c**. Measured radiation heat flow as a function of temperature bias of different gap distances. For the maximum temperature bias considered, the heat flow increases from ~4 nW to ~50 nW as the gap distance decreases from 726 nm to 151 nm, corresponding to maximum radiation heat conductance ~0.26 nW/K. **d**. Heat flow enhancement ratio compared to blackbody radiation calculated by $Q_{bb} = \sigma A F_{12}(T_h^4 - T_s^4)$.

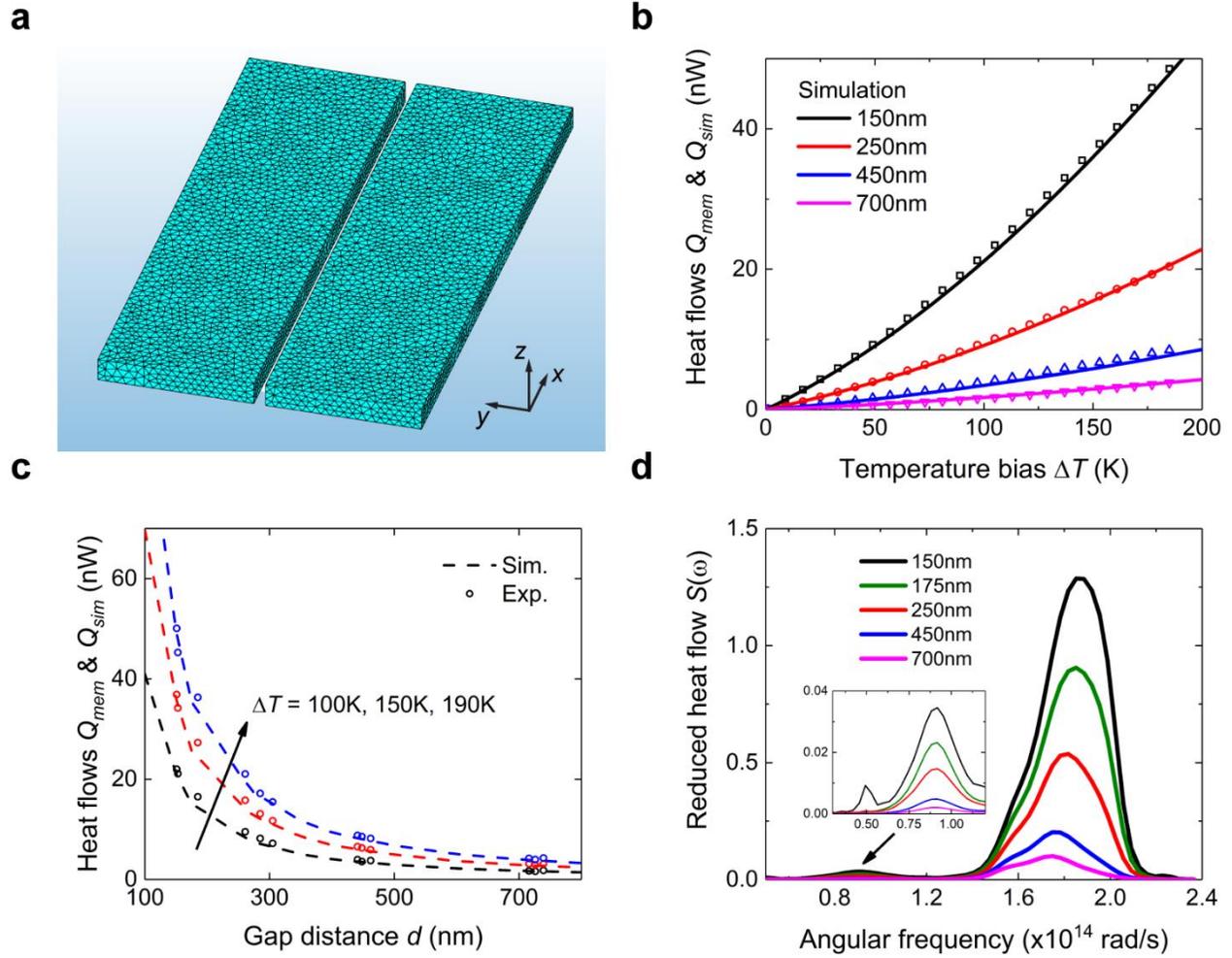

**Figure 3. Fluctuational-EM simulation and comparison with measurements. a.** Surface mesh generated for the FSC method. **b.** Comparison between the simulated and measured heat flows. The measurements show excellent agreement with the simulations. **c**. Comparison between the simulated and measured heat flows as a function of gap distance for the temperature biases of 100, 150 and 190 K, respectively. **d.** Reduced heat flow spectra. The inset shows the zoomed-in spectrum in the low frequency range.

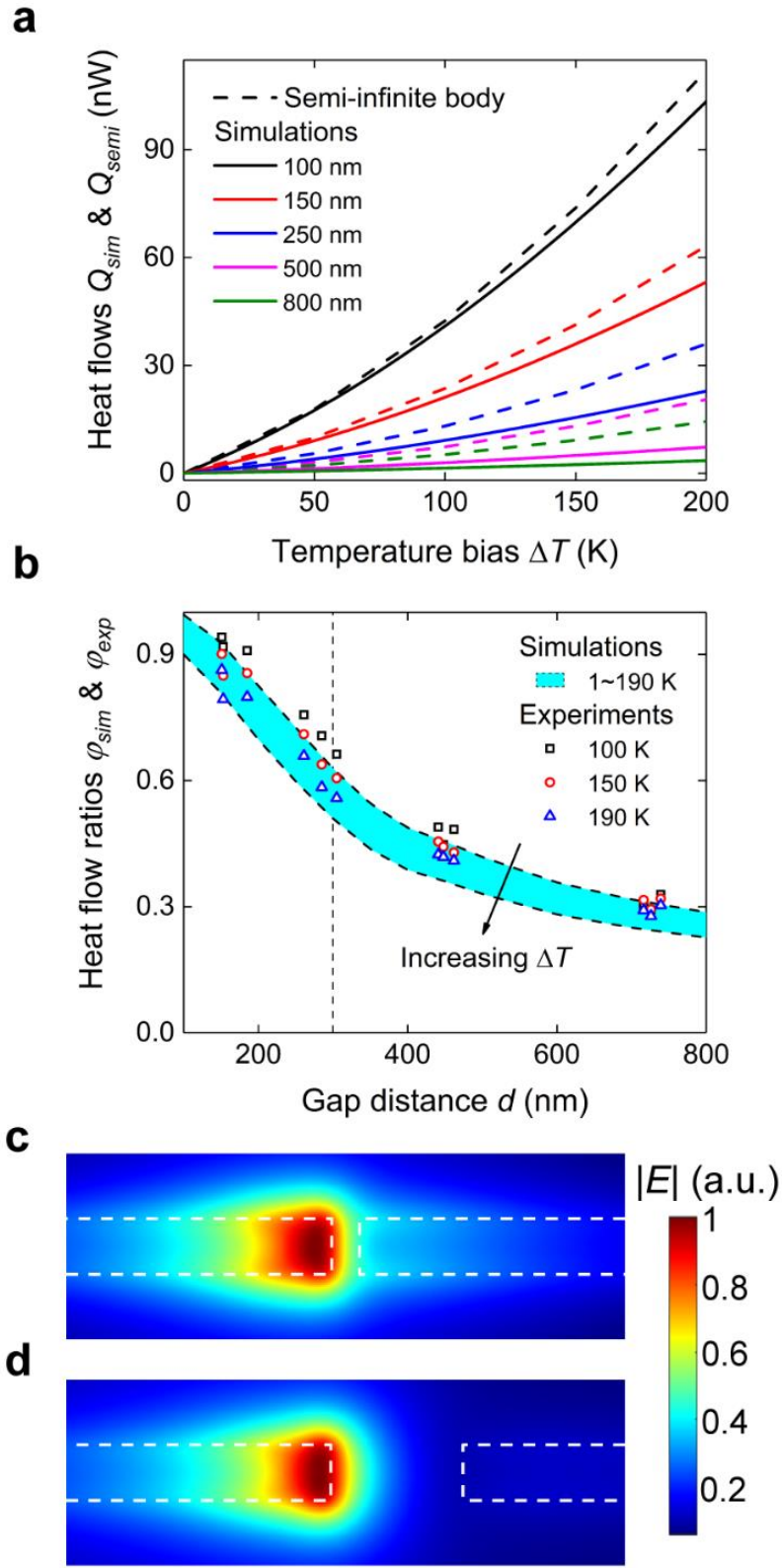

**Figure 4. Theoretical analysis by comparing with the semi-infinite body theory. a.** Comparison between the heat flows from semi-infinite body theory (dashed line) and fluctuational-EM simulation (solid line). The near-field

radiation heat transfer between the sub-wavelength membranes is lower than the prediction between semi-infinite bodies. We attribute the lower heat flow across the sub-wavelength surfaces to the energy carried by the polaritons scattered away from the gap. **b**. Heat flow ratio $\varphi_{sim}$ (shaded region) & $\varphi_{exp}$ (dots) as a function of gap distance, where $\varphi_{sim}$ and $\varphi_{exp}$ are the ratios of fluctuational-EM simulation and experimental measurements to semi-infinite body predictions, respectively. $\varphi_{sim}$ or $\varphi_{exp}$ dramatically increases from ~20% to almost 100% as the gap distance decreases from 800 nm to 100 nm. Besides, $\varphi_{sim}$ or $\varphi_{exp}$ increases with a higher rate as the gap distance decreases below the thickness, 300 nm. **c** & **d**. Electric field profiles between the parallel membranes with gap distances of 150 nm and 700 nm, respectively. The electric field profile at the 150 nm gap distance is more uniformly distributed within the separation.

# Methods

**Fabrication process.** As shown in Extended Data Fig. 1, the fabrication process starts with the deposition of 300 nm thick low-stress silicon nitride films on both sides of a silicon substrate via low-pressure chemical vapor deposition. A 30-nm-thick Pt layer for the heater and sensor structures is then deposited by sputtering and patterned through electron-beam lithography and lift-off processes. A 50-nm Au electrode is subsequently patterned and deposited using another lift-off process. To create the through-hole structure for heat insulation, a silicon-nitride hard mask is formed on the backside of the chip via reactive ion etching. After that, potassium hydroxide (KOH) wet etching is applied to etch most of the silicon beneath the metal structure. The heater and sensor structures of the silicon nitride are formed by electron-beam lithography and reactive ion etching, followed by another KOH wet etching process to suspend the structure. Finally, critical point drying is performed to release the suspended structure. For each batch of devices, there is a reference device for the subtraction of background radiation.

**Characterization of reference devices.** The reference devices closely resemble the devices for near-field thermal measurements. The only difference is that it has a single beam sensor without the absorbing membrane. Extended Data Figs. 2 a & b show the SEM images of a reference device. The purpose of the reference device is to measure any parasitic radiation absorbed by the remaining sensor structure excluding the absorbing membrane. A dedicated reference device is fabricated for each batch of fabrication. The measured parasitic radiation $Q_{ref}$ is subtracted from the experimental heat flows measured for all the near-field devices fabricated in the same batch to find $Q_{mem}$. Extended data Fig. 2c shows the thermal radiation measurement results obtained from all reference devices. To consistently subtract the parasitic radiation, a fifth-order polynomial fitting with least absolute shrinkage and selection operator (LASSO) regularization is applied, as shown by the solid lines in Extended data Fig. 2c.

**Thermal Measurement.** All the near-field thermal radiation measurements are conducted under high vacuum ($10^{-3}$ Pa) and precise temperature control. We first calibrate the heat conductance of the sensor $G_s$ by applying a DC current (up to 20 µA) and measuring the corresponding temperature increase through simultaneously supplying a small sinusoidal current (0.4 µA, 16Hz). Then, we measure the near-field radiation by heating up the emitting membrane using a DC current (up to ~800 µA with a 5 µA interval) and monitor the temperature increase of the sensor membrane using a Wheatstone bridge circuit. Based on the platinum thermometry, we derive the temperature increases from the measured resistance variations (See Supplementary Information for the determination of $\Delta T_h$ and $\Delta T_s$). The calibration of the platinum thermometers includes temperature-resistance calibration, thermal frequency domain and noise floor (See Supplementary Information for the results). We collect ten data points at each DC current applied to the heater and use linear interpolation to determine the heat flow at integer temperature biases (See

Supplementary Information for details). Based on the measured heat flow, we further calculate the radiation heat conductance between the two membranes, $G_{mem} = Q_{mem}/\Delta T$.

**Electric field Simulation.** To quantify the impact of sub-wavelength structures on near-field radiation, we conduct finite-difference time-domain (FDTD) simulations to obtain the electric field profile near the gap. The first 250 nm-deep slot of the emitting membrane is excited based on the Wiener Chaos expansion (WCE) method[42] and discrete dipole approximation. The FDTD simulation is implemented using ANSYS Lumerical. Figures 4 c &d depict the electric field profiles excited by the y-polarized dipoles.

**Acknowledgement**

The authors are grateful to the Defense Threat Reduction Agency (Grant No. HDTRA1-19-1-0028), and the National Science Foundation (Grant No. CBET-1931964). The authors also acknowledge the Claire & John Bertucci Nanotechnology Laboratory at Carnegie Mellon University for the fabrication of the devices and the Material Characterization Facility at Carnegie Mellon University for scanning electron microscopy.

**Author contributions**

X.L. and S.S. conceived the concept of the work. X.L., Z.W., B.Y., and S.D. designed the thermal nanodevice. X.L., Z.W., H.K. and B.Y. developed the fabrication process. X.L. measured the gap distance. H.S. measured the height difference. X.L., Z.W., and H.S. developed the experimental setup and measurement procedure. X.L., Z.L. and J.L. implemented the fluctuational-electromagnetic simulations. H.S. developed the analytical calculation of the near-field radiation between semi-infinite bodies. The manuscript is written by X.L., H.S., and S.S., with comments and input from all authors. X.L., H.S., Z.W., and Z.L. contributed equally. S.S. supervised the research.

**Competing interests**

The authors declare no competing interests.

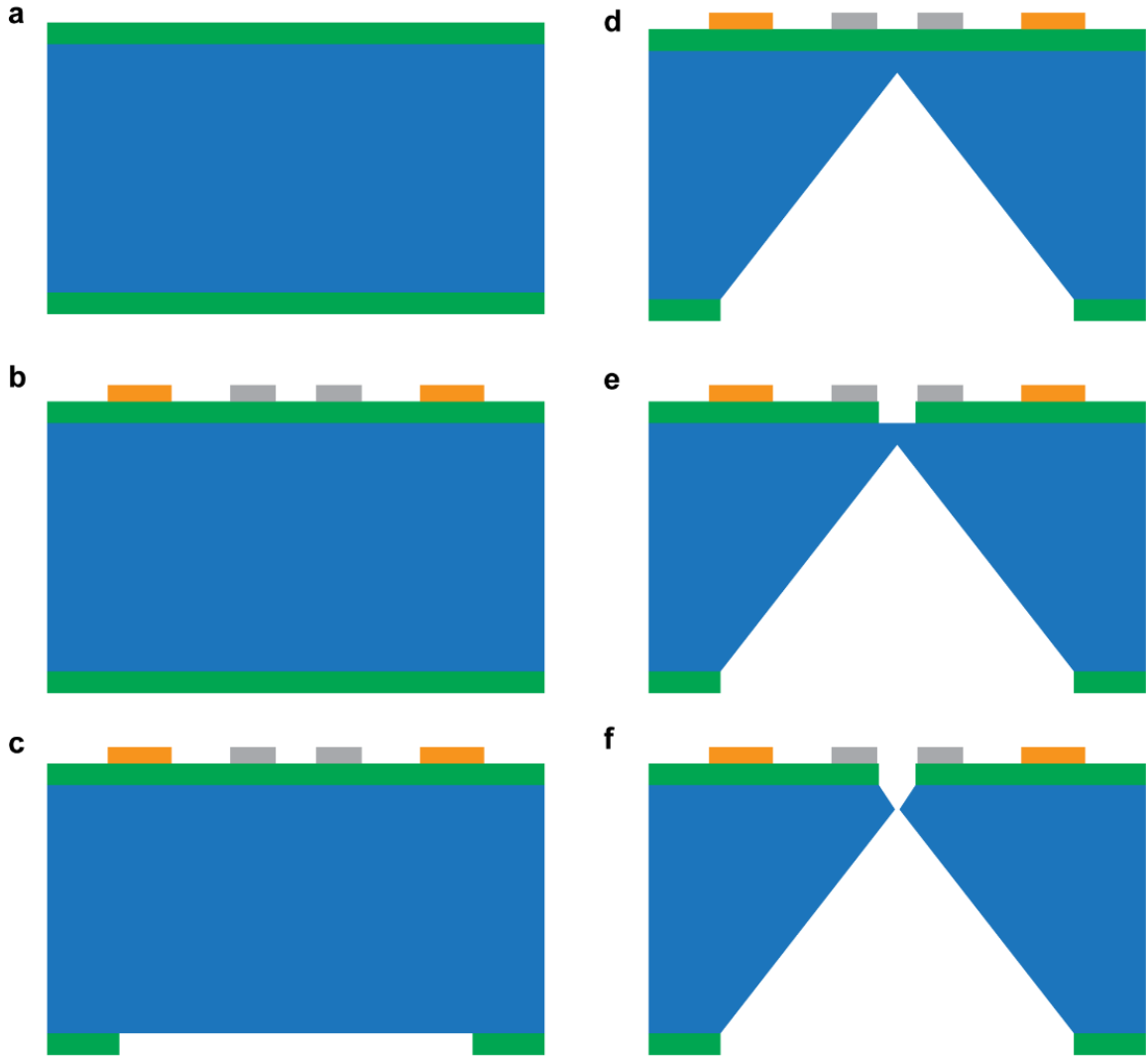

**Extended Data Fig. 1. | Fabrication process flow of the suspended nanodevices. a**, Deposition of low-stress silicon nitride on both sides of the substrate by low-pressure chemical vapor deposition. **b**, Patterning and deposition of the platinum heater and sensor, and gold electrodes. **c**, Patterning and etching of silicon nitride on the backside. **d**, KOH wet etching of silicon. **e**, Patterning and etching of silicon nitride on the frontside. **f**, Second KOH wet etching of silicon and critical point drying.

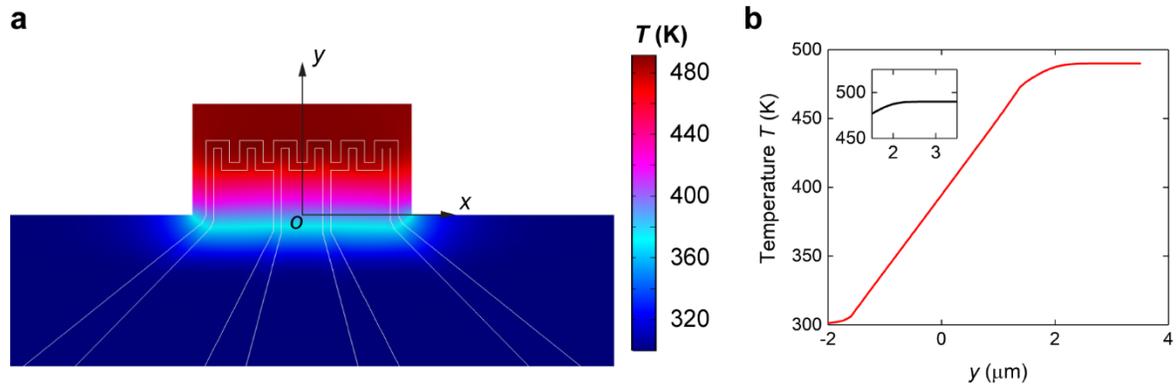

**Extended Data Fig. 2. | Simulated temperature profile on the emitting membrane. a**, Top-view temperature profile of the emitting membrane. Uniform temperature is observed within the first 2-μm of the emitting membrane with the optimized heater structure. The temperature greatly drops beyond the first 2 μm such that the emission from the base is minimized. **b**, Temperature profile on the emitting membrane along the *y* axis. Inset: Temperature profile within the first 2-μm of the emitting membrane.

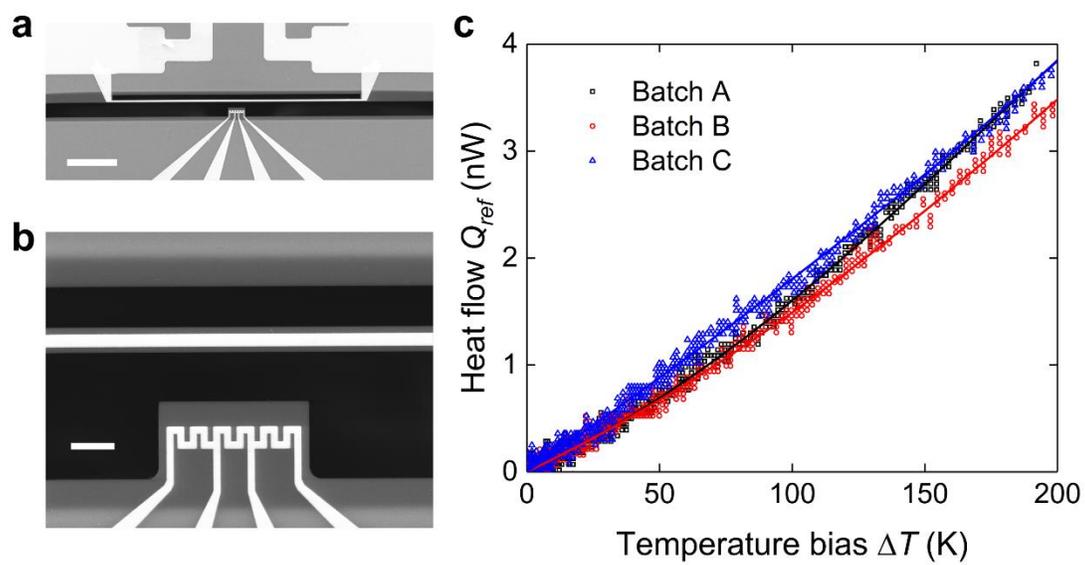

**Extended Data Fig. 3. | Reference nanodevices. a & b**, SEM images of a reference device. Scale bar: 20 μm and 2 μm. **c**, Measured radiation $Q_{ref}$ of reference devices. Solid lines indicate fifth-order polynomial fitting with LASSO regulation.

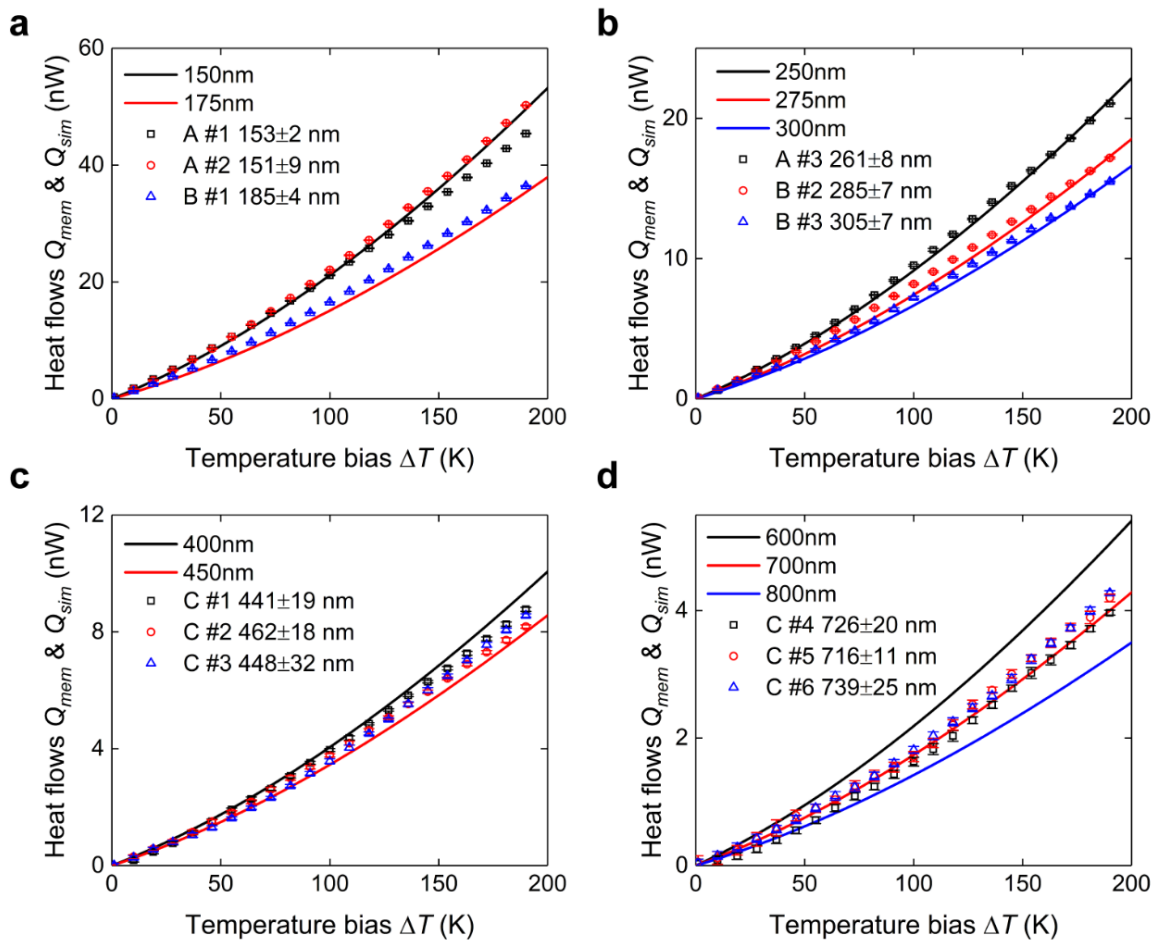

**Extended Data Fig. 4. | Heat flow data of all measured nanodevices.** Markers represent experimental data. Solid lines represent the simulation using FSC method.

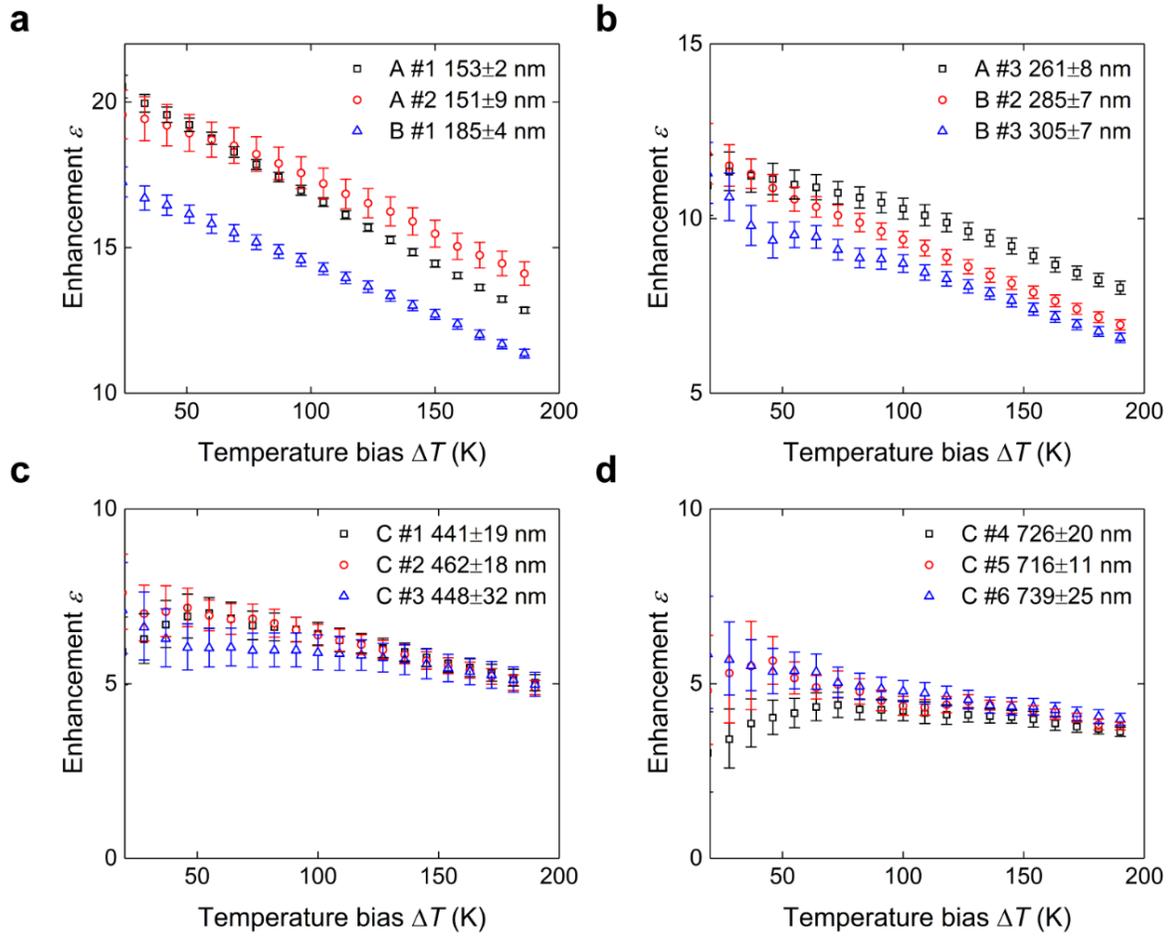

**Extended Data Fig. 5. | Enhancement ratio ε of all measured nanodevices.** For all gap distances, the enhancement ε decreases with higher temperature bias. The error of enhancement ε derives from two factors: the error of interpolated heat flows and the error of view factor from the deviation of gap distance measurements.

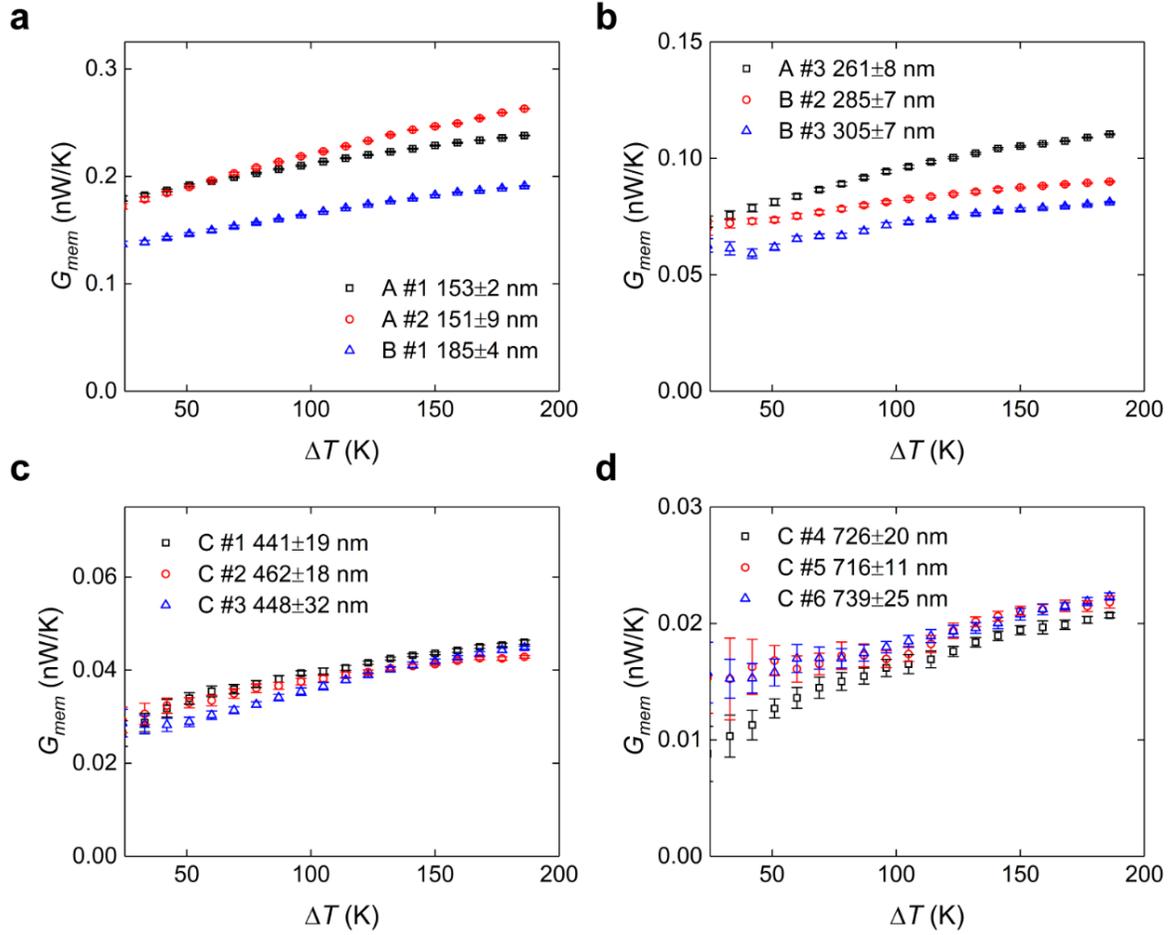

**Extended Data Fig. 6. | Heat conductance of radiation $G_{mem}$ of all measured nanodevices.** For all gap distances, the heat conductance of radiation $G_{mem}$ increases with a higher temperature bias. The error of $G_{mem}$ derives from the heat flow error.

Supplementary Information for

# Nanodevice-Enabled Near-Field Thermal Radiation between Sub-Wavelength Surfaces

Xiao Luo[1,+], Hakan Salihoglu[1,+], Zexiao Wang[1,+], Zhuo Li[1,+], Hyeonggyun Kim[1], Jiayu Li[1], Bowen Yu[1], Shen Du[1], Sheng Shen[1,*].

1 Department of Mechanical Engineering, Carnegie Mellon University, Pittsburgh, PA 15213, USA.

+ The authors contributed equally.

* Corresponding author: Sheng Shen, email: sshen1@cmu.edu

**This PDF file includes:**
    Supplementary Text
    Table S1 Error estimation of the 1D heat transfer model for the sensor.
    Table S2 Gap distances of all tested devices.
    Figure S1 Near-field thermal emission from the emitting membrane.
    Figure S2 Temperature-resistance relation of heater and sensor.
    Figure S3 Thermal frequency domain responses.
    Figure S4 Noise floor of the thermal measurement.
    Figure S5 Linear interpolation of the raw data.
    Figure S6 Finite element simulation of heat transfer within the sensor structure.
    Figure S7 View factor F12 between two parallel 7×0.3 μm² rectangles.
    Figure S8 Dielectric permittivity of the silicon nitride membrane.
    Figure S9 FTIR reflectance measurement of the silicon nitride membrane at different temperatures.
    References

## Near-field thermal emission from the emitting membrane

To investigate the length of the emitting membrane contributing to the near-field thermal radiation, we split the emitting membrane into multiple 7 μm × 0.3 μm × 0.3 μm slots positioned along the length and use the Wiener Chaos expansion (WCE) method[1] to investigate the contribution from each slot to the near-field radiation, by individually exciting each of them, as shown in Fig. S1a. Figure S1b shows the reduced heat flow (with arbitrary unit) spectra of the ten slots from the edge for a gap distance of 150 nm. The magnitude

S1

of the peak in the reduced heat flow drops quickly as the position of the slots gets far from the edge, especially for slots behind slot 7, which corresponds to a distance of ~ 2 μm from the gap.

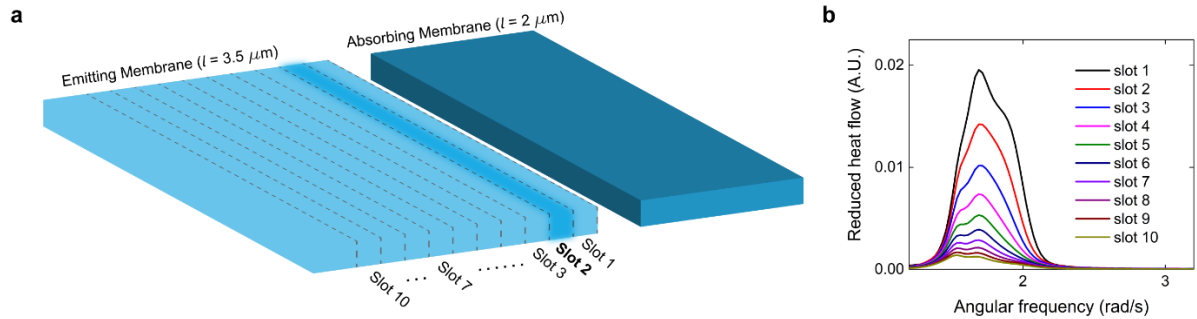

**Figure S1. Near-field thermal emission from the emitting membrane. a,** Schematic of the slot-by-slot Wiener Chaos expansion (WCE) simulation. Each slot, e.g., slot 2 in the schematic, is excited individually. **b,** Reduced heat flow spectra of the ten slots from the edge.

## Temperature-resistance calibration of PRT

The consistent temperature-resistance relation of a platinum resistance thermometer (PRT) enables the determination of the temperature by measuring the resistance. For a small temperature variation (~10K), the linear relation is accurate enough to fit between temperature and resistance. Temperature coefficient of resistance $\alpha_1$ (TCR) of PRT is a measure of the linear relation, which is given by:

$$\alpha_1 = \frac{1}{R_0}\frac{dR}{dT} \tag{S1}$$

where $R_0$ is the resistance at a reference temperature $T_0$. We select 300 K as the reference temperature. For large temperature variations (>100K), the parabolic fitting is more accurate:

$$R(T) = R_0[1 + \alpha_2(T - T_0) + \beta_2(T - T_0)^2] \tag{S2}$$

The temperature-resistance relation calibration of the PRT is performed by sweeping the ambient temperature and measuring the corresponding resistance. Figure S2 shows the measured resistances of the heater and sensor for a temperature range of 280K to 470K. During the near-field radiation measurement, , the sensor undergoes a minor temperature variation and thus linear fitting around 300K is applicable to calibrate the temperature-resistance relation of the sensor. Consequently, the $\alpha_1$ value of the sensor resistance $R_s$ is $1.788\times10^{-3}$ K$^{-1}$. For heater resistance $R_h$, parabolic fitting is applicable, which yields $1.763\times10^{-3}$ K$^{-1}$ for $\alpha_2$ and $-1.046\times10^{-6}$ K$^{-2}$ for $\beta_2$, respectively.

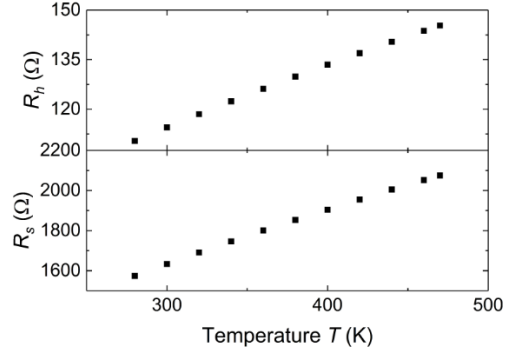

**Figure S2. Temperature-resistance relation of heater and sensor.**

## Thermal frequency domain responses of the heater & sensor

The frequency-domain thermal response is measured by applying a small sine wave current $i$ with a large offset $I_0$ (large enough to induce notable temperature increase) and measure the increase of the sinusoidal voltage amplitude $\Delta V$ using a lock-in amplifier. As suggested by literature[2], the voltage increase $\Delta V$ is a function of the frequency of the sinusoidal current $i$:

$$\Delta V(I) = \begin{cases} 3i\Delta T(I)\dfrac{dR(I=0)}{dT}, f \ll f_0 \\ i\Delta T(I)\dfrac{dR(I=0)}{dT}, f \gg f_0 \end{cases} \tag{S3}$$

where $f_0$ is the thermal cutoff frequency of the heater or sensor. There is a factor-of-3 difference between the voltage increases $\Delta V$ between modulation frequencies much lower and much higher than the thermal cutoff frequency. Figure S3 shows the frequency domain responses of the sensor and heater, respectively. The thermal cutoff frequency of the heater is at ~$10^4$ Hz magnitude, while that of the sensor is ~$10^2$ Hz magnitude.

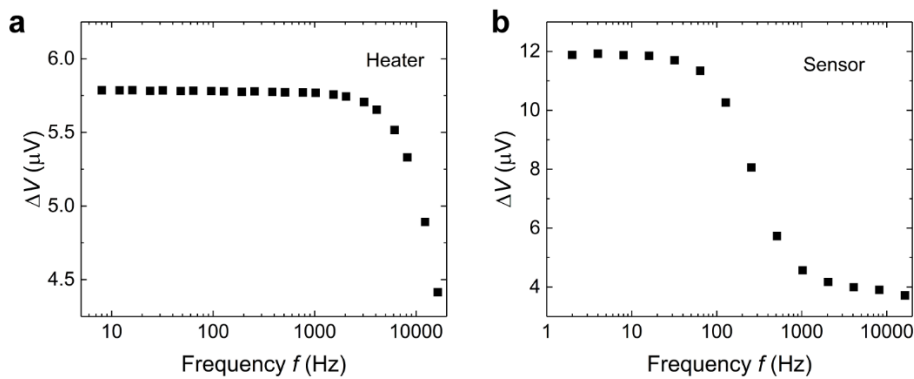

**Figure S3. Thermal frequency domain responses. a**, heater, and **b**, sensor.

# Measurement noise floor

To determine the noise floor of the near-field measurement, we plot the measured heat flow data for a range of small temperature bias. As shown in Fig. S4, the heat flow decreases as the temperature bias decreases from the highest temperature bias in the measurement; however, our measurement technique cannot resolve the heat flows less than ~200pW arising when the temperature bias is below nearly 1 K. Accordingly, the noise floor is determined as ~200 pW.

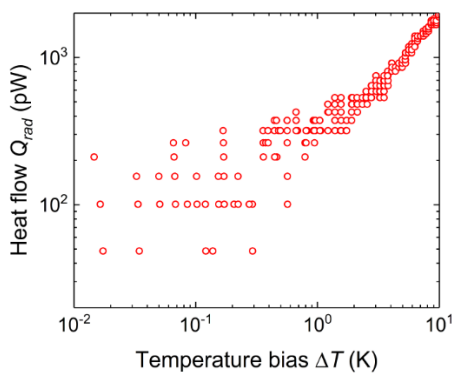

**Figure S4. Noise floor of the thermal measurement.** The noise floor is ~200 pW.

# Linear interpolation of raw data

To condense the measurement data for a given nanodevice, the raw data, e.g., black rectangles in Fig. S5, is linearly interpolated over a range of 16 K at temperature biases with 9-K increments between the neighbor data points, e.g., red filled circles in Fig. S5. Heat flow data reported in Fig. 2 and Fig. 3 in the main text follows this procedure. The maximum error arising from the interpolation is ~0.9 nW.

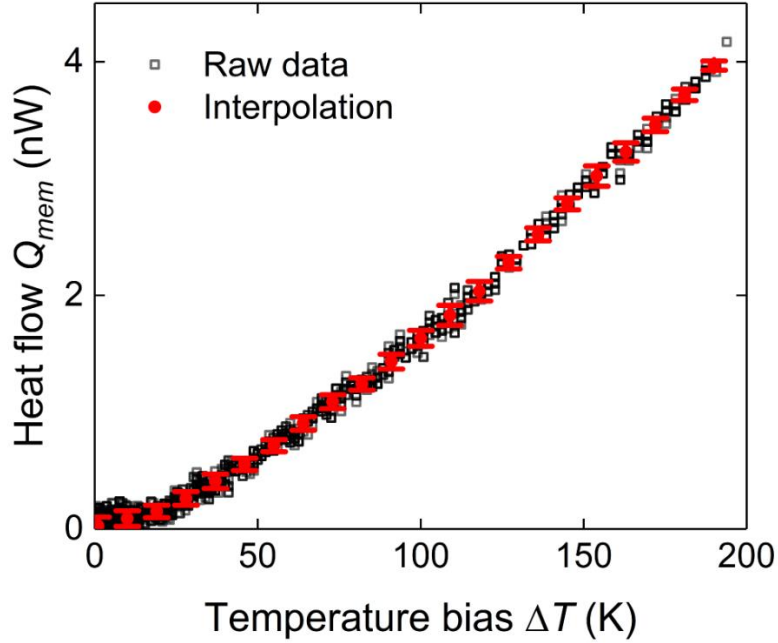

**Figure S5. Linear interpolation of the raw data.** The error bar indicates the error from the linear fitting. The range of interpolation is 16K and the maximum error is ~0.9 nW.

# Determining temperature variations from the platinum thermometer

To determine the temperature variation of the heater, the classical four-probe connection is performed. We use the first harmonic voltage $V_h$ to determine the temperature increase of the heating membrane $\Delta T_h$. The frequency of the sinusoidal current, 211Hz, is much lower than the cutoff frequency of the heater, which indicates the 3 times relation at the low-frequency limit (See Section Thermal frequency domain responses of the heater & sensor). Here, we use a modified low-frequency relation due to the large temperature increase of the heating membrane: $\Delta T_h$ is determined from the first harmonic voltage $V_h$ by solving the equation below,

$$\Delta V_h / i = R_0(\alpha_2 \Delta T_h + \beta_2 \Delta T_h^2) + \frac{2R_0 \Delta T_h (\alpha_2 + 2\beta_2 \Delta T_h)(1 + \alpha_2 \Delta T_h + \beta_2 \Delta T_h^2)}{1 - \beta_2 \Delta T_h^2} \quad (S4)$$

where $i$ is the amplitude of the sinusoidal current.

To calibrate $G_s$ of the sensing beam, the four-probe connection with a low frequency (16Hz) sinusoidal current $i$ is used when applying Joule heating. The 3-time relation at the low-frequency limit is used to determine the spatial-average temperature increase of the sensor $\overline{\Delta T_{s,J}}$ due to Joule heating,

$$\overline{\Delta T_{s,J}} = \frac{\Delta R_{s,J}}{3\alpha_1 R_{s,0}} \tag{S5}$$

where $\Delta R_{s,J} = \Delta V_{s,J}/i$, and $R_{s,0}$ is the resistance of the sensor at 300K.

During the near-field radiation measurement, the Wheatstone circuit is used to determine the temperature increase of the sensor. A sinusoidal 0.01V voltage is applied to the Wheatstone bridge, and the frequency is optimized to 1531 Hz for minimizing the influence of self-heating. The spatial-average temperature increase of the sensor can be determined from the resistance change,

$$\overline{\Delta T_s} = \frac{\Delta R_s}{\alpha_1 R_{s,0}} \tag{S6}$$

## Heat transfer within the sensor beam and calibration of $G_s$

The heat transfer along the sensor beam can be approximated by a 1D heat conduction model. We approximate the near-field heat flow as a localized heat input at the center of the beam sensor. Also, we assume the temperature at the base of the sensor equals the ambient temperature, then the heat transfer equation is,

$$-kA\frac{d^2T}{dx^2} = 0 \tag{S7a}$$

$$T(x = L) = T_{amb}, -kA\left.\frac{dT}{dx}\right|_{x=0^+} = \frac{1}{2}Q_{rad} \tag{S7b}$$

$$T(x = -L) = T_{amb}, -kA\left.\frac{dT}{dx}\right|_{x=0^-} = \frac{1}{2}Q_{rad} \tag{S7c}$$

where $T_{amb}$ is the ambient temperature, and the $x$ coordinate spans from -L to L. A is $(A_{Pt} + A_{SiN})$. The effective thermal conductivity $k = (k_{Pt}A_{Pt} + k_{SiN}A_{SiN})/(A_{Pt} + A_{SiN})$. By applying the temperature boundary conditions (S7b, c) to Eq. S7a, the temperature distribution is derived as

$$T(x) = \begin{cases} -\dfrac{Q_{rad}}{2kA}x + \dfrac{Q_{rad}}{2kA}L + T_{amb}, x \geq 0 \\ \dfrac{Q_{rad}}{2kA}x + \dfrac{Q_{rad}}{2kA}L + T_{amb}, x < 0 \end{cases} \quad (S8)$$

Then, the spatial-average temperature increase of the sensor $\overline{\Delta T_s}$ is derived by integrating along the sensor,

$$\overline{\Delta T_s} = \frac{1}{2L}\int_{-L}^{L} T(x)dx - T_{amb} = \frac{Q_{rad}}{4kA}L \quad (S9)$$

Here, the temperature of the absorbing membrane $T_s$ is equal to the temperature at the center ($x = 0$), and the heat conductance of the sensor $G_s$ is,

$$G_s \equiv \frac{Q_{rad}}{\Delta T_s} = \frac{Q_{rad}}{2\overline{\Delta T_s}} = \frac{2kA}{L} \quad (S10)$$

where $\Delta T_s$ is the temperature increase of the absorbing membrane ($x = 0$).

To calibrate the heat conductance of the sensor beam $G_s$, we apply a heating current $I$ and measure the corresponding average temperature increase $\overline{\Delta T_{s,J}}$, induced by the Joule heating power $Q_h$. The corresponding heat transfer problem can be modelled by

$$-kA\frac{d^2T}{dx^2} = \frac{Q_h}{2L} \quad (S11a)$$

$$T(x = L) = T(x = -L) = T_{amb} \quad (S11b)$$

$$Q_h = I^2 R_s \quad (S11c)$$

By solving Eq. S11 with the given boundary conditions, the temperature distribution is found as

$$T(x) = -\frac{Q_h}{4kAL}x^2 + \frac{Q_h L}{4kA} + T_{amb} \quad (S12)$$

Then the spatial-average temperature increase of the sensor due to Joule heating $\overline{\Delta T_{s,J}}$ is,

$$\overline{\Delta T_{s,J}} = \frac{1}{2L}\int_{-L}^{L} T(x)dx - T_{amb} = \frac{Q_h}{6kA}L \quad (S13)$$

Then, by substituting the $kA/L$ in Eq. S10 with Eq. S13, we obtain

$$G_s = \frac{1}{3}\frac{Q_h}{\overline{\Delta T_{s,J}}} \quad (S14)$$

To test the accuracy of the 1D heat transfer model, we simulate the heat transfer in the sensor structure using finite element simulation, implemented by COMSOL Multiphysics. As shown in Fig. S6a, we first simulate the Joule heating scenario by applying a uniform heat source with 1 µW total power and then calculate the heat conductance *Gs* using Eq. S14. We then simulate the near-field radiation scenario by applying a 1 µW power heat input on the 7×0.3 µm² lateral surface of the absorbing membrane, as shown in Fig. S6b. Figure S6c shows the temperature profile along the sensor of the two cases: radiation heat input indicates an almost linear temperature distribution while Joule heating indicates a parabolic temperature distribution. To validate the heat transfer model for the sensor, we calculate the radiation heat flow by $Q_{rad} = 2G_s \overline{\Delta T_s}$ by using the simulated temperature distribution and compare it with the real heat flow - 1 µW, as shown in Table S1. The error rate is less than 1%. To further estimate the error of thermal radiation to the ambient, we assume blackbody emissions on all the surfaces of the sensor and redo the procedure above. The error rate increases but is still less than 1%.

**Table S1. Error estimation of the 1D heat transfer model for the sensor.**

| Ambient radiation | $\overline{\Delta T_s}$ (K) | $\overline{\Delta T_{sJ}}$ (K) | $G_s$ (nW/K) | Q (nW) real | Q (nW) calculated | Error (%) |
|---|---|---|---|---|---|---|
| N | 8.2300 | 5.4998 | 60.610 | 1000 | 997.6 | -0.24 |
| Y | 8.1084 | 5.4276 | 61.415 | 1000 | 995.9 | -0.41 |

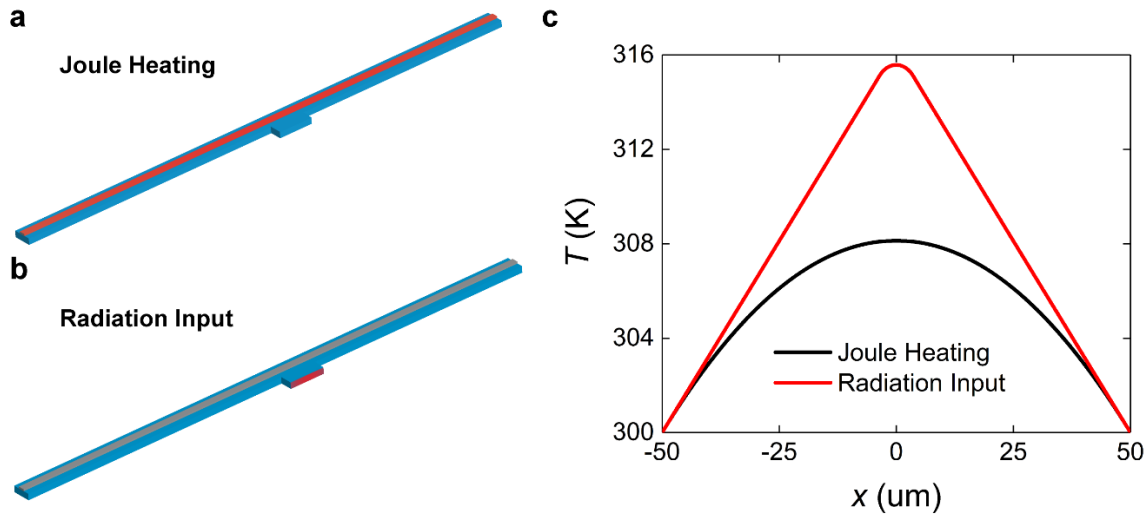

**Figure S6. Finite element simulation of heat transfer within the sensor structure. a, & b,** the two simulation scenarios, Joule heating and radiation input, for determining the error of the 1D heat transfer model. **c,** The temperature profiles along the sensor of these two scenarios.

# Gap distances measurements of the devices

We take top-view SEM images of the devices and use the valley width of the grayscale profile to determine the gap distances. For each device we took at least three images with different magnifications. The mean value and standard deviation are used as gap distances and errors, respectively. The gap distances are listed in Table S2.

**Table S2. Gap distances of all tested devices.** * indicates the device whose results are presented in Fig. 2 and Fig. 3. Measured radiation heat flows of all devices are available in Extended Data.

| Batch | Device | Gap Distance (nm) |
|---|---|---|
| A | #1 | 153±2 |
| | #2* | 151±9 |
| | #3* | 261±8 |
| | #4 | 2133±5 (ref) |
| B | #1 | 185±4 |
| | #2 | 285±7 |
| | #3 | 305±7 |
| | #4 | 2285±6 (ref) |
| C | #1* | 441±19 |
| | #2 | 462±18 |
| | #3 | 448±32 |
| | #4* | 726±20 |
| | #5 | 716±11 |
| | #6 | 739±25 |
| | #7 | 2467±9 (ref) |

# View factor calculation

The view factor between the exchanging surfaces is calculated by[3]

$$F_{12} = \frac{2}{\pi XY} \left( ln \left\{ \left[ \frac{(1+X^2)(1+Y^2)}{1+X^2+Y^2} \right]^{\frac{1}{2}} \right\} + X\sqrt{1+Y^2} \, tan^{-1} \frac{X}{\sqrt{1+Y^2}} + Y\sqrt{1+X^2} \, tan^{-1} \frac{Y}{\sqrt{1+X^2}} - X \, tan^{-1} X - Y \, tan^{-1} Y \right) \quad (S15)$$

where $X = \frac{t}{d}, Y = \frac{w}{d}$. $t$, $w$, and $d$ are the thickness and width of the membrane, and gap distance, respectively.

Figure S7 shows the view factors between two 7×0.3 µm² surfaces as a function of gap distance.

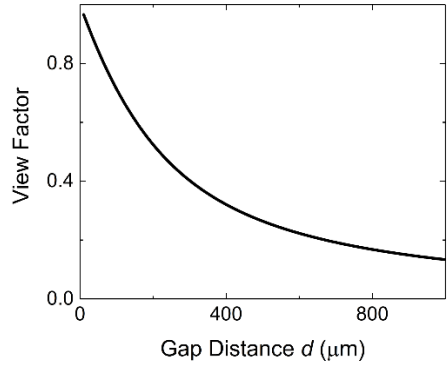

**Figure S7. View factor $F_{12}$ between two parallel 7×0.3 μm² rectangles.** The view factor drastically decreases as the gap distance increases from 100 nm to 1 μm.

# Dielectric permittivity of silicon nitride

We use Maxwell-Helmholtz-Drude dispersion model[4] to characterize the dielectric properties of the silicon nitride membranes used in our experiments, as plotted in Fig. S8.

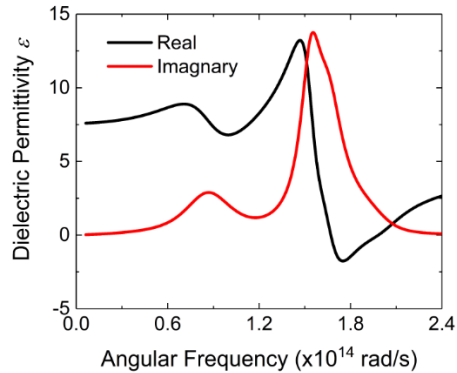

**Figure S8. Dielectric permittivity of the silicon nitride membrane.**

# Reflection measurement of the silicon nitride membrane

To determine the influence of temperature to the dielectric properties of silicon nitride, we measure the reflectance of suspended silicon nitride thin films for a temperature range of 323K to 498K, using Fourier Transfer Infrared Spectroscopy (FTIR). As shown in Fig. S9, the measured reflectance shows a robust reflectance profile, especially near resonance frequency (1000 cm$^{-1}$), and thus we assume that the dielectric properties remain constant in the temperature range considered.

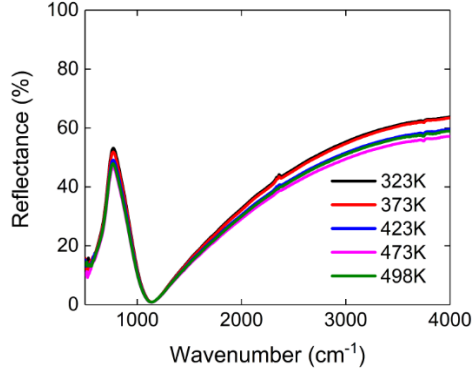

**Figure S9. FTIR reflectance measurement of the silicon nitride membrane at different temperatures.**

## Semi-infinite body calculation

We consider two planar continuous media, each at uniform hot, $T_H$, and cold, $T_C$, temperatures and separated by a vacuum gap, $d$, along the $z$ axis. The heat flux from the hot (1) to cold medium (2) reads[5]:

$$Q'' = \int \frac{d\omega}{2\pi}[\Theta(\omega, T_1) - \Theta(\omega, T_2)] \int \frac{d\kappa}{2\pi} \kappa \times \mathcal{T}_q \qquad (S16)$$

$\Theta(\omega, T)$ is mean energy of Planck oscillators with frequency $\omega$ at temperature $T$ and is given by $\hbar\omega/(e^{\hbar\omega/k_B T} - 1)$, where $\hbar$ and $k_B$ are the reduced Planck constant and Boltzmann constant, respectively. $\kappa$ denotes the wave vector component parallel to planar surface ($\kappa = k_x \hat{x} + k_y \hat{y}$). $\mathcal{T}_q$ represents energy exchange function and is given as $\frac{(1-|r_q|^2)(1-|r_q|^2)}{|1-r_q r_q e^{2ik_{z,o}d}|^2}$ for propagating, $\kappa < \omega/c$, and $\frac{4Im(r_q)Im(r_q)e^{-2Im(k_{z,o})d}}{|1-r_q r_q e^{2ik_{z,o}d}|^2}$ for evanescent, $\kappa > \omega/c$, waves. $r_q$ stands for complex Fresnel reflection coefficient ($r_q = Re(r_q) + iIm(r_q)$ where $i$ is imaginary unit) for q-polarization ($q = s, p$). Here, $r_s = \frac{k_{z,o}-k_{z,m}}{k_{z,o}+k_{z,m}}$ and $r_p = \frac{\varepsilon k_{z,o}-k_{z,m}}{\varepsilon k_{z,o}+k_{z,m}}$. Also, $k_{z,o} = [\omega^2/c^2 - \kappa^2]^{1/2}$, and $k_{z,m} = [\varepsilon\omega^2/c^2 - \kappa^2]^{1/2}$ are dispersion relations of polaritons in free-space and materials, respectively, where $c$ is the speed of light, and $\varepsilon$ is the dielectric permittivity values of the low-stress silicon nitride. To find heat transfer rate, $Q_{semi}$, reported in Fig. 4a, we exploit $Q_{semi} = Q'' \times A$ where $A$ stands for the cross-section area ($7 \times 0.3$ μm$^2$).

## Polariton and wavelength relation

At the separation regimes of interest, $\kappa$ has approximately the same magnitude of $Im(k_{z,o})$ from the dispersion relation ($\kappa \gg \omega/c$). We know that wavenumber is inversely proportional to wavelength ($\kappa = 2\pi/\lambda$ where $\lambda$ is wavelength). This relation can be given for both lateral axes, separately. At relatively large distances within the interested regime, the polaritons contributing to the integration over $\kappa$ in Eq. S16 have small $\kappa$ values due to $e^{-2Im(k_{z,o})d}$ term in energy exchange function in the near field. Thus, these polaritons have large wavelengths. At shorter distances, the contributing polaritons have large $\kappa$, possessing small wavelengths.